# Gibbs energy of ices III, V and VI: wholistic thermodynamics and elasticity of the water phase diagram to 2300 MPa.

B. Journaux*[1], J.M. Brown[1], A. Pakhomova[2], I. E. Collings[3,4], S. Petitgirard[5], P. Espinoza[1], J. Ott[1,6], F. Cova[3,7], G. Garbarino[3], M. Hanfland[3]

## 1. Abstract:

Gibbs energy representations for ice III, V and VI are reported. These were constructed using new measurements of volumes at high pressure over a range of low temperatures combined with calculated vibrational energies grounded in statistical physics. The collection of representations including ice Ih and water (released as the open source "SeaFreeze" framework) allow accurate determinations of thermodynamics properties (phase boundaries, density, heat capacity, bulk modulus, thermal expansivity, chemical potentials) and seismic wave velocities over the entire range of conditions encountered in hydrospheres in our solar system (220-500K to 2300 MPa). These comprehensive representations allow exploration of the rich spectrum of thermodynamic behavior in the $H_2O$ system. Although the results are broadly applicable in science and engineering, their use in habitability analysis in water-rich planetary bodies of our solar system and beyond is particularly relevant.

## 2. Introduction

Water is a fundamentally important molecule in scientific fields ranging from biology to engineering, earth and environmental sciences, chemistry or astrophysics. As a common molecular species in our cosmic neighborhood, water ice polymorphs at high pressures in planetary interiors could be the most abundant "mineral group" in the Universe (Hanslmeier 2011). A focus on potentially habitable icy moons and ocean exoplanets hydrospheres (Sotin and Tobie 2004; Vance and Brown 2013; Journaux et al. 2013; Noack et al. 2016; Journaux et al. 2017) motivates an interest in thermodynamic properties of water and ices in the < 2 GPa range. For example, the presence of an insulating layer of high-pressure ice between the deep ocean and the underlying silicates on large water-rich planetary bodies has been identified as a potential bottleneck for habitability as it would limit nutrient transport (Léger et al. 2004; Noack et al. 2016; Journaux et al. 2017). Thus, accurate thermodynamic representations for all stable phases of water and aqueous solutions are essential in the analysis of potential planetary habitability.

[1]University of Washington, Department of Earth and Space Sciences, Seattle, USA. [2]Photon Sciences, Deutsches Elektronen-Synchrotron, Hamburg, Germany. [3]European Synchrotron Radiation Facility, Grenoble, France, [4]Empa - Swiss Federal Laboratories for Materials Science and Technology, Center for X-ray Analytics, Dübendorf, Zürich, Switzerland [5]ETH Zurich, Institute für Geochemie und Petrologie, Zürich, Switzerland. [6]University of California Santa Cruz, Department of Earth & Planetary Sciences, Santa Cruz, USA. [7]Norwegian University of Science and Technology, Department of Physic, Trondheim, Norway.

Unfortunately, sparse measurements at high pressure, some performed more than 100 years ago, limit our understanding of high-pressure ice thermodynamics and phase equilibria. In order to better constrain the structure, evolution and habitability of the interiors of water-rich planetary bodies, a new generation of accurate measurements and internally-consistent thermodynamic representations of aqueous solutions and ice polymorphs is required. Furthermore, since next-generation planetary exploration missions are likely to investigate the seismic structure of icy worlds (Vance et al. 2018), data on seismic wave speeds in ices and aqueous solutions as a function of pressure and temperature are also needed.

Bridgman (1912) provided the first delimitation of the phase boundaries for ices Ih, II, III, V and VI. These measurements and additional work for the ices VI-VII phase boundaries (Bridgman 1937) constitute the main (and sometimes only) constraints for the water phase diagram boundaries below 2 GPa (V. Tchijov et al. 2004; M Choukroun and Grasset 2007; Dunaeva, Antsyshkin, and Kuskov 2010; Wagner et al. 2011). Bridgman provided numerous pressure–temperature points along the phase boundaries as well as volume changes measured using the displacement of the piston of his high-pressure apparatus. Crystallographic structures and absolute volume measurements of ice polymorphs came later as X-ray and neutron diffraction measurements were obtained on metastable (quenched) samples retrieved cryogenically (McFarlan 1936a, 1936b; Barclay Kamb and Davis 1964; Barclay Kamb 1965; B. Kamb, Prakash, and Knobler 1967). Surprisingly few additional measurements have been reported for these ices since the pioneering work.

Current thermodynamic models for the water phase diagram at higher pressures (V. Tchijov et al. 2004; M Choukroun and Grasset 2007; Mathieu Choukroun and Grasset 2010; Dunaeva, Antsyshkin, and Kuskov 2010), use *ad hoc* parameterizations to independently compute chemical potentials, volumes, and specific heats of each phase. Since all equilibrium thermodynamic quantities (e.g. volume, chemical potential, bulk modulus, thermal expansivity, etc.) are derived from an underlying thermodynamic potential (*eg*. Gibbs energy), properties should not be independently parameterized. While these models were adjusted to reproduce observed phase equilibria and volume changes by Bridgman (1912), they remain thermodynamically inconsistent which excludes derivation of other related thermodynamic properties. Extrapolations of properties beyond the range of experimental constraints are not reliable and the substantial effort necessary to implement these representations remains a barrier to their widespread use. A comprehensive approach using the Gibbs energy (as a function of pressure and temperature) has been reported for ice Ih (Feistel and Wagner 2006). The main obstacle to the extension of that effort to high-pressure ice polymorphs III, V, and VI has been the lack of adequate in situ high-pressure measurements that extend below room temperature.

Volume measurements at high pressure using neutron diffraction with $D_2O$ ice III were reported by Londono, Kuhs, and Finney (1998) (3 data points at 245-250 K) and Lobban, Finney, and Kuhs (2000) (5 data points in the 250-330 MPa and 240-250 K range). Two piston-displacement points for ice III at 248 K are given by Shaw (1986). Gagnon et al. (1990) gave an isothermal (238 K) equation fitted to four unreported data points which unfortunately provides unrealistic volumes as pointed out by Choukroun and Grasset (2007).

Ice V appears to have been barely more studied *in situ* with two piston-displacement points (248 K) from Shaw (1986) and a density-pressure law from Gagnon et al. (1990), with the latter again providing unreasonable volumes, as also noted above for ice III. $D_2O$ ice V neutron diffraction volumes were collected over a wider range and represent to this day the largest dataset on ice V with 11 points in the 400-500 MPa and 100-254 K range (Lobban, Finney, and Kuhs 2000).

The first extensive pressure-temperature-volume data for $H_2O$ ice VI were published by Bezacier et al. (2014) but only above 300 K. A review of other datasets for $H_2O$ and $D_2O$ ice VI measurements is also provided in Bezacier et al. (2014).

To provide the requisite data for construction of better thermodynamic representations, we obtained new X-ray diffraction measurements on ice III, ice V, and ice VI in a pressure regime from 200 to 2000 MPa for temperatures between 220 and 270 K. Mie-Grüneisen equations of states were determined, where pressure is separated into the cold (zero Kelvin) compression plus a thermal pressure calculated using quasi-harmonic vibrational energies based on physically motivated phonon densities of states that are constrained by measurements and theory. The resulting representation of Gibbs energy for each of the ices and for liquid water (Bollengier et al., 2019), using the framework described in Brown (2018), then provides a fundamental and robust description of their thermodynamic properties. Phase equilibria (solid-solid and solid-liquid) are accurately determined as the locus of pressure-temperature points with equal chemical potential and equilibrium thermodynamic properties (e.g, density, specific heat, bulk modulus, thermal expansivity, entropy and enthalpy) are derived from appropriate analytic derivatives of Gibbs energy (e.g, density, specific heat, bulk modulus, thermal expansivity, entropy and enthalpy). In addition, isotropic seismic wave velocities are estimated for all polymorphs as a function of pressure and temperature using a volume and temperature dependent shear modulus that is constrained by the ultrasonic and Brillouin measurements at high pressure combined with the adiabatic bulk modulus from the Gibbs energy representation.

This water and ices "toolkit", provided as the "SeaFreeze" framework in supplementary materials, allows a self-consistent exploration of thermodynamic properties and elasticity of water and ices over a large range of pressure-temperature conditions covering all planetary hydrospheres of our solar system. An anticipated extension of SeaFreeze will also include solute thermodynamics.

## 3. Results:

### 3.1 New X-Ray diffraction measurements

High-pressure low-temperature X-ray diffraction measurements of ice polymorphs were performed at the ID15B beamline (ESRF, France) using cryo-cooled diamond anvil cell. The flat-panel Mar555 area detector was used. Ice powders of the forms III, V, and VI were obtained by freezing supercooled liquid water far from the melting line. Several volumes were obtained on unstrained single crystals formed by annealing near the melting point. Volumes as a function of pressure were measured on several isotherms for each polymorph (Table 1). All powder and single crystal patterns could be fitted with the space groups $P4_12_12$, $C2/c$, and $P4_2/nmc$ for ice III, V and VI respectively. More details on the experimental methods can be found in the methods section and examples of diffractogram refinement are given in Supplementary materials.

### 3.2 Thermodynamic Representations for ice polymorphs:
#### 3.2.1 *Gibbs energy equation of state*

Analytic local basis function (LBF) representations of Gibbs energy (Brown 2018) using Mie-Grüneisen equations of state for each ice polymorph were constructed using (i) our new measurements of volumes as a function of pressure and temperature, (ii) thermal pressures based on quasi-harmonic phonon (vibrational) densities of state, (iii) reported elastic moduli measured as a function of pressure and temperature, (iv) estimates of the configurational entropy $S_o$ and Gibbs energy $G_o$ at the reference state, and (v) previously determined liquid-solid phase boundaries (i.e. melting curves). The supporting equations and fitting procedures are described in the methods sections. Table 2 provides sources and data associated with the quasi-harmonic phonon densities of states and Table 3 lists equation of state parameters ($K_o$, $K_o'$, $V_o$ and $\gamma_o$). Relevant constraints for $G_o$ and $S_o$ for each ice polymorph and the value in our models are reported in Table 4

In Figure 1 pressures are shown as a function of volume for the three high-pressure ices. Based on the Mie-Grüneisen approach, measurements (filled symbols) minus calculated thermal (vibrational) pressures define "corrected" points (open symbols) that should lie within measurement uncertainty of the cold compression curve (dashed lines parameterized using $K_o$, $K_o'$ and $V_o$). The 1-bar volumes for ice V and ice VI quenched to 98.15 K (Kamb 1965) (largest volume data for each ice) have relatively small thermal pressures and thus provide robust estimates of $V_o$. The quenched volume reported by Kamb for ice III is inconsistent with the current measurements and was excluded from the fit. For ice VI, our new cryogenic measurements at high pressure were combined with 45 points previously reported in Bezacier et al. (2014) ranging from 1260 to 2560 MPa and from 300 to 340 K and measured using the same experimental setup and procedures. The combined data covers the entire range of pressure stability of this polymorph. The thermodynamic Grüneisen parameters, $\gamma_o$, were chosen to be consistent with spectroscopically-determined mode $\gamma$'s (Table 2) and to reduce misfit of data on the cold compression curve. A systematic trend of larger $\gamma_o$ values for higher pressure ices ranging from 1 to 1.4 is apparent in Table 2 and Table 3.

The pressure derivative of the isothermal bulk modulus, $K_o'$, is poorly constrained by measurements that span a relatively small range of compression. A value of 4 was assumed in the previous work of Fortes et al. (2012) and Bezacier et al. (2014). However, a strong constraint on the pressure dependence of the adiabatic bulk modulus is provided by ultrasonic and Brillouin measurements (Gagnon et al. 1988, 1990; Tulk, Kiefte, et al. 1997; Tulk, Gagnon, et al. 1997; Shimizu et al. 1996; Shaw 1986). Simultaneous fits to compressions (**Figure 1**) and the adiabatic moduli (**Figure 2a)** require that $K_o'$ range from 6 for ice III and ice V to 6.5 for ice VI.

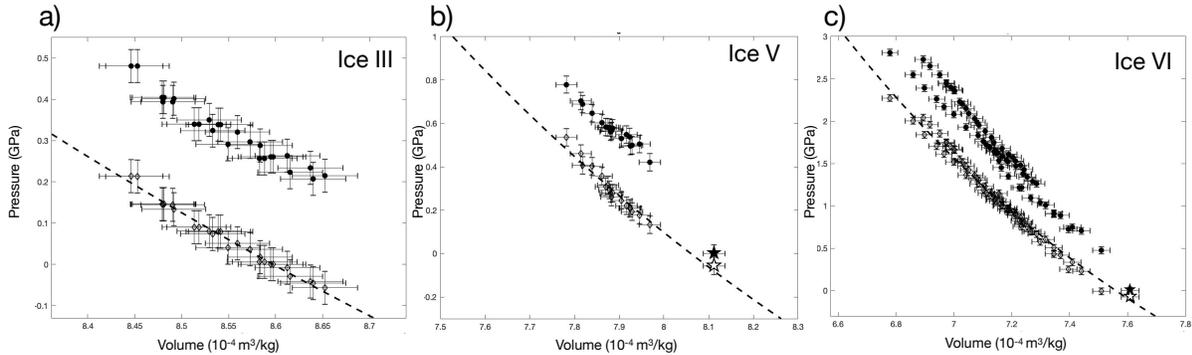

Figure 1: Pressures as a function of specific volume for a) ice III, b) ice V and c) ice VI. Filled symbols are measurements. Open symbols are measurements minus $P_{therm}$ that provide an estimate of the zero-Kelvin compression curve. Stars for ice V and ice VI are the ambient pressure volumes from Kamb (1965). Experimental uncertainties are indicted as 30 MPa in pressure and 0.3% in volume. The dashed curves are the result fits of the cold-compression (zero-Kelvin) data.

### 3.2.2 *Shear modulus and seismic velocities*

The isotropic elastic shear modulus, $\mu$, determined as the mean of Hashin-Shtrikman bounds (Brown 2015) for single crystal measurements or directly from transverse wave speed measurements in isotropic polycrystalline samples (Gagnon et al. 1988, 1990; Tulk, Kiefte, et al. 1997; Tulk, Gagnon, et al. 1997; Shimizu et al. 1996; Shaw 1986), are also shown in **figure 2a**. A simple linear function in density and temperature ($\mu = A + B*\rho + C*T$) with the same parameters adequately fits the shear moduli of all three high pressure ices while different parameters were required for ice Ih (see details in **Supplementary materials**). Using this parameterization and the adiabatic bulk modulus and density from the Gibbs energy representations, isotropic body wave velocities (compressional and transverse) are compared as a function of pressure with measurements in **Figure 2b**. The computed sound speed match most of data within a few percent with the exception of the shear wave velocities reported for ice V by Shaw (1986).

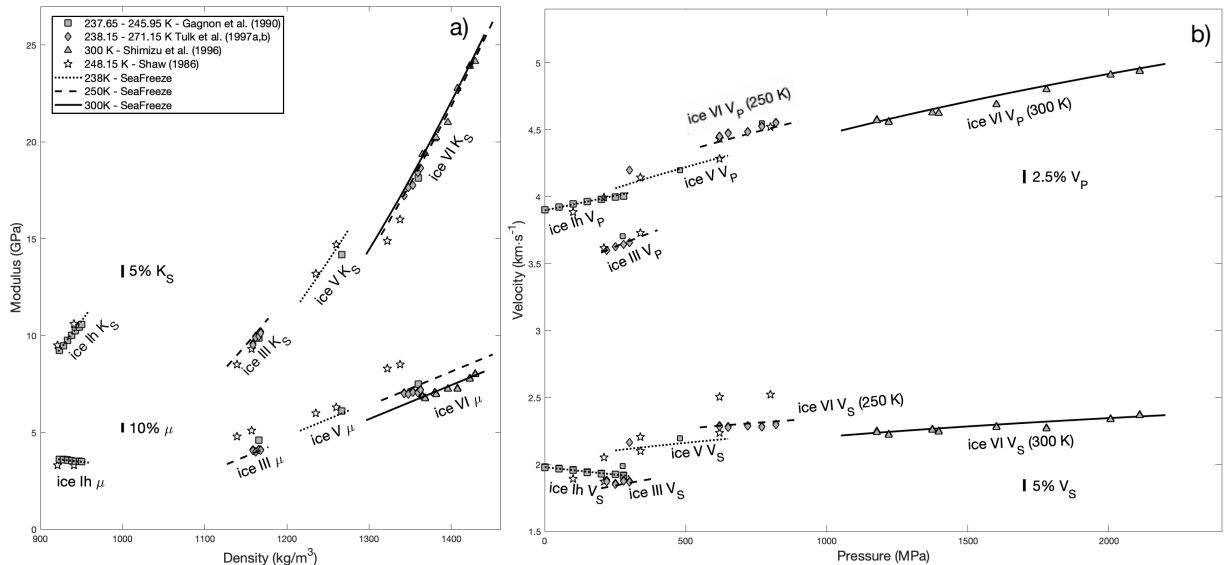

Figure 2: Isotropic aggregate elastic properties for ice Ih, III, V, and VI. a) bulk $K_s$ and shear $\mu$ moduli as a function of density, and b) isotropic P and S-waves velocities as a function of pressure. Different symbols (described in the legend) are used for datasets from different studies. Brillouin and sound speed data (Gagnon et al. 1988, 1990; Tulk, Kiefte, et al. 1997; Tulk, Gagnon, et al. 1997; Shimizu et al. 1996; Shaw 1986). Solid (300 K), dashed (250 K) and dotted (238 K) lines are predictions based on the SeaFreeze representations. Vertical scales are indicative of data scatter apparent in the figures.

3.2.3 *Ice polymorph melting curves*

The predicted phase boundaries for both solid-liquid and solid-solid transitions from the intersections of Gibbs surfaces are shown in **Figure 3** along with experimental determinations. Temperature residuals of data from the predicted melting points are plotted in **Figure 4**. The Gibbs energy for Ice Ih is a direct LBF parametrization of the Feistel and Wagner (2006) equation of state.

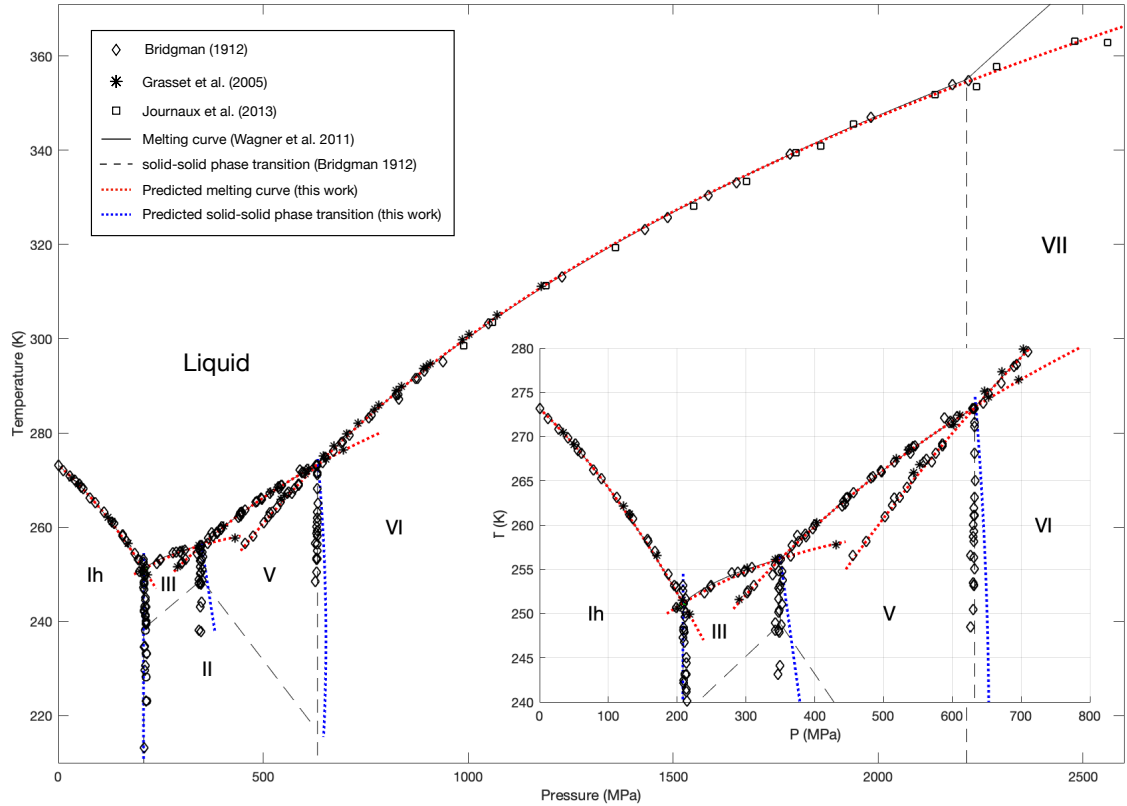

Figure 3: Water phase diagram. Ices polymorphs melting curves and solid-solid phase transition calculated using the Gibbs LBF representations in red and blue dotted lines respectively. Melting curves from the Simon-Glatzel equations for the melting curves from Wagner et al. (2011) represented with a thin black line, often overlap by the LBF predicted melting lines at this scale. Solid-solid phase transition from Bridgman (1912) are represented as dashed black lines. Experimental data are represented as open diamond from Bridgman (1912) and Bridgman (1937), black asterisks from Grasset, Amiguet, and Choukroun (2005), open squares from Journaux et al. (2013) and open circles for Kell and Whalley (1968). A zoom on the ice Ih-III-V melting region is represented on the bottom right.

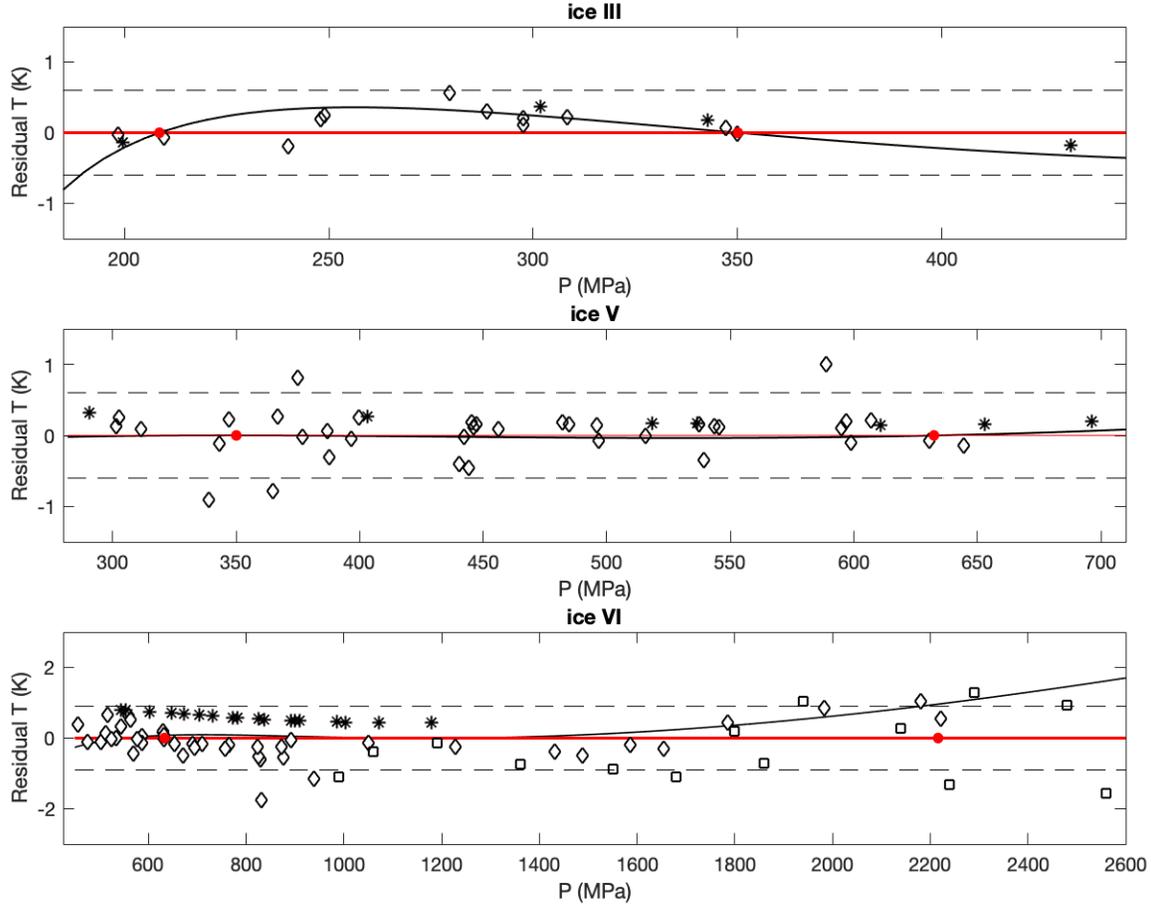

Figure 4: Temperature residuals of the melting curves data (see figure 3 for symbols references) and the Simon-Glatzel equations from Wagner et al. (2011) as black lines, with our LBF calculated melting boundaries (red line). Propagated temperature uncertainties of 0.6 K for ice III and ice V and 0.9 K for ice VI (corresponding to the reported pressure and temperature uncertainties) are represented as black dashed line. Triple points from Wagner et al. (2011) are represented as red dots.

More than 90% of the reported melting points (including metastable determinations) lie within measurement uncertainty of the Gibb energy-determined melting lines. The current melting curve for ice V is indistinguishable from the Simon-Glatzel parametrization of Wagner et al. (2011). Although the ice III melting curve of Wagner et al. (2011) has larger curvature than the current prediction, both lie within the 0.6 K uncertainty range. Measurements of ice VI melting are well represented over a much larger range of pressure than for the other lower pressure polymorphs. The clear curvature of the melting line shown in Figure 3 is well matched by the current representations, lending credence to the underlying physical model. Previously reported locations of the VI-VII-liquid triple point range from 352.2 to 355K and from 2160 to 2216 MPa (Wagner et al. 2011; Journaux et al. 2013). Using Wagner et al. (2011) Simon-Glatzel parametrization for the ice VII melting curve, our estimated VI-VII-L triple point at 353.5 K and 2200 MPa lies within the prior bounds.

### 3.2.4 *New prediction of ices solid-solid phase transition*

The solid-solid phase transitions are model predictions since no optimization was undertaken to match these transition. As shown in Figure 3, the predicted ice Ih – ice III transition matches measurements while the predicted ice III – ice V and ice V – ice VI transitions are systematically offset by up to 30 MPa. The Ih-III transition was determined (Bridgman 1912; Kell and Whalley 1968) in reversed measurements (crossing the boundary during both compression and decompression). In contrast, the determinations for ice III – ice V and ice V – ice VI by Bridgman (1912) represent pressures of transition during decompression only. Based on normal hysteresis of solid-solid transitions requiring significant structural reorganizations, it is likely that such measurements would significantly underestimate the transition pressure. For example, Pistorius, Rapoport, and Clark 1968 reported 100 MPa of hysteresis for the ice VI – VII transition. We suggest that our estimates for the solid-solid transitions better represent thermodynamic equilibrium than determinations based on the unreversed measurements.

### 3.2.5 *Metastable range and equilibrium thermodynamic properties*

The current Gibbs energy representations are expected to sensibly predict properties both within and beyond the stability range of each ice phase. In **Figure 5,** an illustration of thermodynamic properties determination for ice V is presented over a wide range of temperatures and pressures. Other polymorphs thermodynamic surfaces are provided in supplementary materials. All surfaces extrapolate with reasonable trends into the metastable regions and have correct limiting behavior at absolute zero where thermal expansivity and specific heat go to zero. Since the characteristic temperatures of molecular vibrational modes (libration, bending and stretching) are so high, specific heat remains temperature dependent and, in the stability field of this phase, is approximately half the high-temperature Dulong-Petit limit (4157.2 J/kg/K). The only other specific heat estimations for high pressure ice derived by Tchijov (2004), following on the ice VII calculations of Fei, Mao, and Hemley (1993), are up to 35% higher than the current analysis. Their use of the Debye approximation (inappropriate for molecular solids) does not adequately account for the large number of high frequency molecular modes.

### 3.2.6 *The "SeaFreeze" thermodynamic framework*

The LBF Gibbs surfaces of ice III, V and VI are combined with the liquid water representation of Bollengier et al. (2019), and the representation for ice Ih (Feistel and Wagner 2006) (converted to the LBF format) as the open-source computational tool "SeaFreeze" (in Python and Matlab™) that is provided in supplementary materials. The SeaFreeze representations are thermodynamically consistent within and between phases. Gibbs energies and entropies of all phases are referenced to IAPWS-95 values for water at its vapor-fluid-ice Ih triple point (Bollengier et al., 2019). This tool gives equilibrium thermodynamic properties (including phase boundaries, density, heat capacities, bulk modulus, thermal expansivity, chemical potentials) as well as shear moduli and seismic wave velocities for each phase extending into metastable regimes.

Example of the evolution of major thermodynamic properties and seismic velocities for all ices are provided in figure 5.b on several isotherms for each ice polymorphs.

The SeaFreeze implementation is computationally efficient. More than $10^5$ thermodynamic points per second can be determined on a mid-range 2015-vintage laptop computer (2.7GHz Intel core i5-5257U CPU). Thus, use of SeaFreeze within computationally intensive frameworks (eg. geodynamic simulations) appears possible.

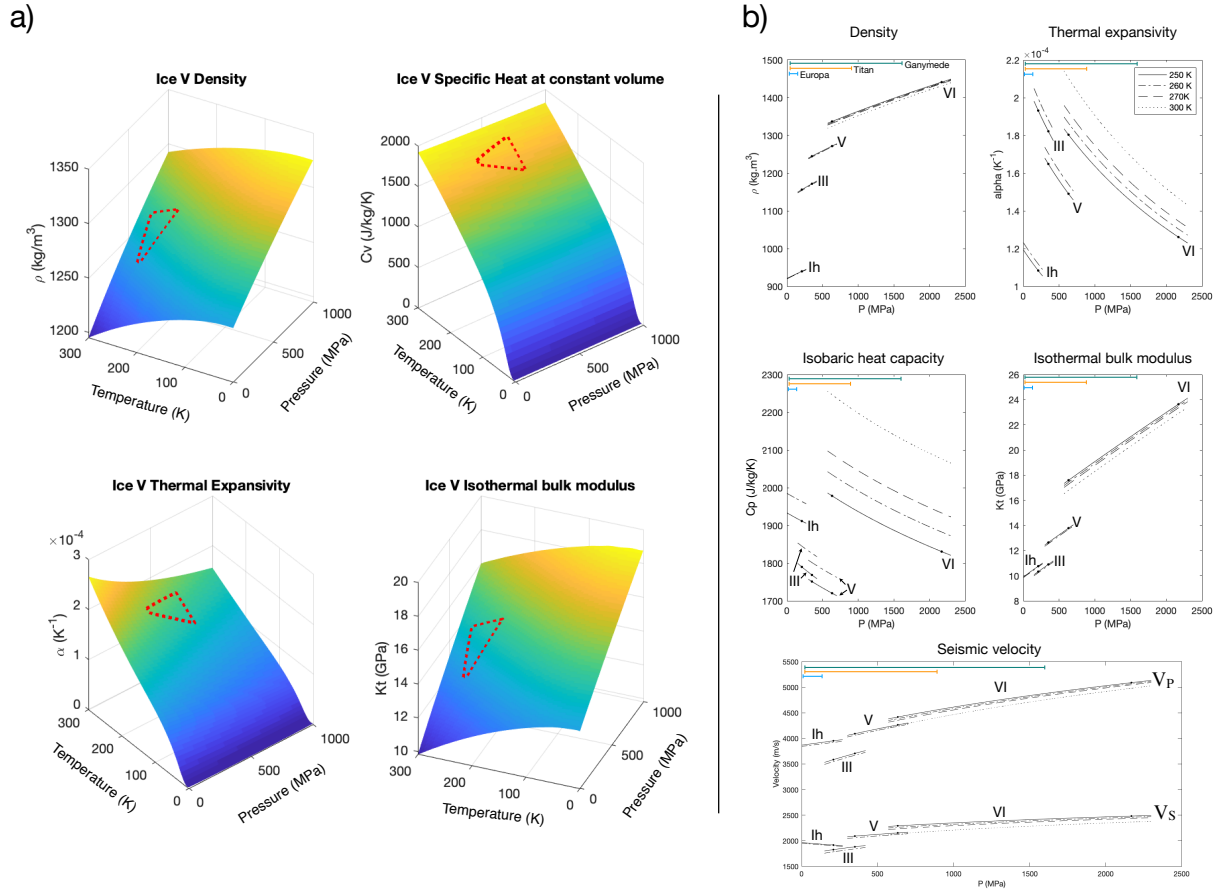

Figure 5: a) Thermodynamic properties of ice V determined as derivatives of the LBF representation of Gibbs energy inside and beyond its stability range (red dashed lines). b) Example of thermodynamic properties and seismic velocities obtain using SeaFreeze. The estimated pressure range found in the hydrosphere of Europa, Titan and Ganymede are reported as blue, yellow and green bars respectively (S. D. Vance et al. 2018). Black dots on the 250K profile represent the solid-solid phase transition.

## Discussion

A uniform numerical environment for the thermodynamic properties of water and phases in equilibrium with water from ambient to high pressures supports research in a broad range of fields including astrobiology, planetary sciences, oceanic science, geochemistry, and condensed matter physics. Studies, previously impeded by disjoint and thermodynamically inconsistent parameterizations that provided inaccurate estimates, can be more readily undertaken within the new framework. SeaFreeze offers clarity in how measurements inform the representations and consistency in how properties relate to energy and entropy reference states. The transparency of the process provides an open-source opportunity to improve and extend the framework. Other phases can be included (eg. ice II and additional stable or metastable ice polymorphs), as well as solid phases of interest for planetary hydrospheres (salts and hydrates). Representations for electrolyte solutions are under development. Not included at this time in SeaFreeze are transport properties such as thermal and electrical conductivities, viscosities of fluids and rheological properties of solids. These could find a place within the SeaFreeze framework.

In planetary sciences, the scientific objectives of upcoming flagship missions (NASA Europa Clipper (Phillips and Pappalardo 2014) and ESA/JUICE (Grasset et al. 2013)) include a strong emphasis on the study of water-rich planetary interiors and the potential habitability of the icy moons. Instruments planned for these spacecrafts will determine properties (e.g. gravitational moments, induced magnetization) that are dependent on the high-pressure aqueous system thermodynamics. Prior studies of interiors and geodynamics of icy moons and water-rich planets (*eg*. Noack et al. 2016; Choblet et al. 2017; Kalousová et al. 2018; Marounina and Rogers 2019) have relied on *ad hoc* thermodynamic parameterizations with tenuous physical basis and based on limited data. The present framework provides an accurate, physically self-consistent description of the candidate constituents that will significantly aid investigations of the icy moon interiors and of their potential habitability. The recent success of the Mars InSight seismometer on the Martian surface highlights the importance of seismic investigations to determine planetary structure. Two concept lander missions for Europa (Europa Lander (Pappalardo et al. 2013)) and Titan (New Frontier Dragonfly mission (Lorenz et al. 2018)) include a seismometer package. Since SeaFreeze body wave velocities in ices and fluids can be implemented in seismic models, this framework can be used to refine seismic instrument specifications and ultimately be a resource in the analysis of mission data.

In summary, using state of the art techniques, we accurately measured long-needed crystal volumes of ice III, V and VI at cryogenic temperatures over a large pressure and temperature range within and beyond their stability fields. By combining these data with a physically grounded thermal model to determine the Gibbs energy, we provide the first equation of state for ice III and V, and significantly extend and enhance previous ice VI representations (Bezacier et al. 2014). These Gibbs representation allows accurate determination of melting curves (including metastable extensions) to 2300 MPa. Moreover, our vibrational energies, based on a statistical physics model and experimental and computational densities of state for each polymorph, allow us to derive for the first time physically grounded specific heat for ice III, V and VI. Our comprehensive Gibbs representation refine the ice III-ice V and ice V-ice VI phase boundaries, recognizing the hysteresis in the unreversed measurements of Bridgman (1912). We combined these Gibbs representation in an open-source and modular computational framework "SeaFreeze" (in Python and Matlab™) providing access to thermodynamic properties of water and ices under conditions found in icy moon hydrospheres.

## Methods
### X-Ray measurements

Ultrapure MiliQ™ water was loaded in 500 to 600 μm culets diamond anvil cell (DAC), equipped with an indented 80 μm thick stainless-steel gasket, with a 300 μm pressure chamber. Ruby was used as a pressure calibrant, with a precision of 30 MPa, using an external ruby in contact with the back of the anvil to correct for the temperature dependence of the fluorescence. Temperature was regulated using a He-cryostat. Diffraction data were collected at the ID15B beamline of the European Synchrotron Radiation Facility (ESRF), Grenoble, France, using a wavelength of 0.41137 Å and a beam diameter of 10 x 10 μm. Diffraction images were collected using a MAR555 flat panel detector. The detector to sample distance was calibrated with a silicon standard using the procedure implemented in Fit2D. The program Dioptas (Prescher and Prakapenka 2015) was used for masking diamond peaks and integrating the 2D images into 1D powder diffraction patterns. Powder diffraction data were collected by a continuous ω rotation of ±5° with a 2 s exposure time. In the case of single crystals, a continuous ω rotation of ±20° with 2 s exposures was used. Refinement of the powder diffraction patterns were performed using Topas (Coelho 2018).

Ice III was formed by cooling water down to 220 K at 0.3 GPa by by 1 K/min, in the stability field of ice II. Data measured directly after freezing at 220K showed texturing and strain upon formation with no clear volume-pressure dependence and were therefore not considered in the fit. Upon heating, relaxation was observed, along with the formation of larger crystal domains, and clear volume trend with temperature could be measured. Both single crystal data (step scans +/-25° every 0.5°) and single frame images (continuous +/-20° rotation scan) were measured and integrated in CrysAlisPro or in Dioptas, respectively. Both refinements were included in the fit.

Ice V was crystallized in its stability field by cooling liquid water at 0.56 GPa down to 230 K. Upon heating and decompression, at ~ 0.45 GPa and 245K the appearance of single reflections of an unknown phase has been observed. The new phase could not be identified with known water polymorph structures and will be the subject of future studies. Ice V and the new phase coexisted over the range of ~220-260K and 0.45-0.55 GPa. Upon heating, the ice V underwent progressive texturing, indicated by substitution of the powder-like rings by sharp diffraction peaks. Several P-T points were excluded from the fit due to the texturing of ice V and/or dominated presence of the unknown phase.

The polycrystalline ice VI was obtained by compressing ice V above 0.65 GPa at 220K. As for ice III, data collected at 220 K directly after the solid-solid transformation shown no clear volume-pressure dependence and were therefore not considered in the fit. Upon heating, relaxation occurred and clear volume trend could be measured up to 262 K.

As represented in figure S4 in supplementary, some data were acquired beyond their stability range taking advantage of the large metastability of ice. Single crystals were formed at the melting curve to ensure pressure and temperature measurements accuracy.

**Gibbs Energy surfaces**

Gibbs energies for ices III, V, and VI at all relevant pressures and temperatures are numerically evaluated through successive integrations starting at reference values of $G_o$ and $S_o$ where the subscript denotes absolute zero and ambient pressure.

$$G(P,T) = \int_{P_o}^{P} V(P,T)dP + G(P_o,T) \quad (1)$$

A Mie-Grüneisen equation of state (described below) is used to represent $V(P,T)$ and the temperature dependence of Gibbs energy at the reference pressure is given as

$$G(P_o,T) = G_o(P_o,T_o) - S_o(P_o,T_o) \cdot (T-T_o) + \int_{T_o}^{T} C_P dT - T \int_{T_o}^{T} \frac{C_P}{T} dT \quad (2)$$

The chosen values for $G_o$ and $S_o$ and for the (adjustable) parameters associated with the Mie-Grüneisen equation of state are discussed in the following section. The constant pressure specific heat, $C_p$, is determined from the constant volume specific heat, $C_v$, that is based on the quasi-harmonic phonon energy and the derivatives of $V(P,T)$ that establish the isothermal bulk modulus, $K_T$, and thermal expansivity, α:

$$C_P = C_V(P,T) + V(P,T) \cdot T \cdot K_T(P,T) \cdot \alpha^2(P,T) \quad (3)$$

Local basis function representations of *G(P,T)* are then determined by collocation (details provided in Supplementary materials). Using the LBF Gibbs energy representations for each phase, solid-liquid and solid-solid phase boundaries are determined as the locus of pressure-temperature points with equal chemical potentials, and thermodynamic properties are analytically calculated from appropriate derivatives of Gibbs energy.

**Mie-Grüneisen Equation of State:**

The Mie-Grüneisen equation of state permits a physically motivated representation, valid over a wide range of pressures and temperatures, that is implemented using a small number of adjustable parameters. Total pressure is separated into a static compression component (cold compression curve at 0 K) and a thermal component associated with lattice vibrations:

$$P(V,T) = P_{0K}(V) + P_{therm}(V,T) \qquad (4)$$

where the cold compression curve $P_{0K}$ is described using 3rd order Eulerian finite-strain formalism:

$$P_{0K} = 3K_o f_E (1 + 2f_E)^{5/2} \left(1 + \frac{3}{2}(K'_o - 4)f_E\right) \qquad (5)$$

with the Eulerian strain given as $f_E = [(V_o/V)^{2/3} - 1]/2$, $K_o$ and $K_o'$ are the bulk modulus and its pressure derivative at 1bar and 0 K.

The thermal pressure is:

$$P_{therm} = \frac{\gamma}{V} \cdot E_{vib}(T,V) \qquad (6)$$

where $\gamma$ is the thermodynamic Grüneisen parameter and $E_{vib}(T,V)$, is the lattice vibrational energy. A power-law volume dependence is added to the Grüneisen parameter:

$$\gamma(V) = \gamma_0 \cdot \left(\frac{V}{V_0}\right)^q \qquad (7)$$

Although values for $q$ between 1 and 2 are commonly used (Brown 1999), here for a restricted regime of compressions, a temperature-independent Grüneisen parameter with $q = 1$ is assumed.

**Vibrational Energy model**

The quasi-harmonic energy $E_{vib}(T,V)$ is computed using a phonon density of states (DoS) for each ice polymorphs where $g(\nu, V)$ gives the number of modes lying in a frequency range between $\nu$ and $\nu + d\nu$. Ice, as a molecular solid, has nine degrees of freedom per molecule that naturally correlate with four families of modes, each with a distinct range of frequencies. These are translational modes (0-500 cm⁻¹), librational modes (500-1100 cm⁻¹), bending modes (1500-1800cm⁻¹) and stretching modes (3200-3800 cm⁻¹) with respectively 3, 3, 1 and 2 degrees of freedom. We use a combination of inelastic neutron scattering (INS) determinations for translational and librational modes of ice V and VI, and computed density of states where INS results are not available. A summary of the sources used in constructing the density of states, $g_{i,0}(\nu, V_0)$, for the four families at 0 K and 1 bar is presented in **Table 2**. As the INS data were obtained on quenched samples at cryogenic temperatures and since thermal expansion tends to zero at absolute zero, these DoS determinations are associated with the reference volume $V_o$.

For constant mode Grüneisen parameters, $\gamma_i = \frac{d \ln \nu_i}{d \ln V}$, the volume dependences of phonon frequencies follow as:

$$\nu_i(V) = \nu_i(V_o) \left(\frac{V_o}{V}\right)^{\gamma_i} \quad (8)$$

which allows computation of the volume dependence of $g_i(\nu, V)$. Although modes within families of modes can have a different $\gamma_i$, here a single value (listed in **Table 2**) is assumed for each of the four vibrational families. Limited data for ices and other solids (based primarily on Raman and infrared measurements) indicate broad trends with translational modes showing values in a range from 1 to 2. In contrast, the internal stretching and bending bonds show little or negative changes in frequency with compression.

For ice V and VI, translational and stretching modes $\gamma_i$ are estimated from experimental in-situ spectroscopic studies of Raman-active mode shifts with pressure (Minceva-Sukarova, Sherman, and Wilkinson 1984). Libration and bending are more challenging to observe in-situ in DACs as they are masked by the diamond's first and second order Raman signal. As no experimental nor computational data exist for those, we chose to use the same approach as for ice Ih, where the $E_{vib}$ is already known in details (Feistel and Wagner 2006). For ice Ih translation and libration from Li (1996) and computed bending and stretching from Jenkins and Morrison (2001) give a value of $\gamma_i$ of 1 for libration and 0 for bending modes. This provides a description with less than 5% error for $E_{vib}$. For Ice III we used the mean Grüneisen associated with the computed DoS family (Ramírez et al. 2012).

The integral of each DoS mode family is normalized by its degrees of freedom $\mathcal{N}_i$, such that $\int g_i \, d\nu = \mathcal{N}_i$, and assembled into the final DoS as a sum of all mode families $g(\nu, V)$. The quasi-harmonic energy $E_{vib}$ per molecule of H$_2$O is then computed as a function of temperature and volume:

$$E_{vib}(T, V) = \int \frac{h\nu}{\exp(h\nu/kT) - 1} \cdot g(\nu, V) \, d\nu \quad (9)$$

where $k$ is Boltzmann's constant, $h$ is Plank's constant, and the integral is over the full frequency range of the vibrational modes.

The isochoric heat capacity is obtained as the temperature derivative of the vibrational energy at constant volume:

$$C_v = \left.\frac{dE_{vib}}{dT}\right|_V \quad (10)$$

An example of a constructed DoS as a function of volume and resulting $E_{vib}$ and $C_v$ for ice V is shown in **figure 6**. The density of state for each ice is reported in Supplementary materials.

In the range of temperatures associated with the equilibrium stability of these ices (<355 K) only translational and librational modes are sufficiently populated and contribute significantly to thermal properties. As necessary by construction, the model asymptotes the Dulong-Petit limit for water (4157.2 J/kg/K) at very high temperatures (**figure 6**). However, in the range of stability of these ices the $C_v$ remains highly temperature dependent. Although explicit anharmonic contributions to the vibrational energy are likely at high temperatures, they are generally thought to be small in comparison with the harmonic contributions at modest or low temperatures and are ignored in the current analysis.

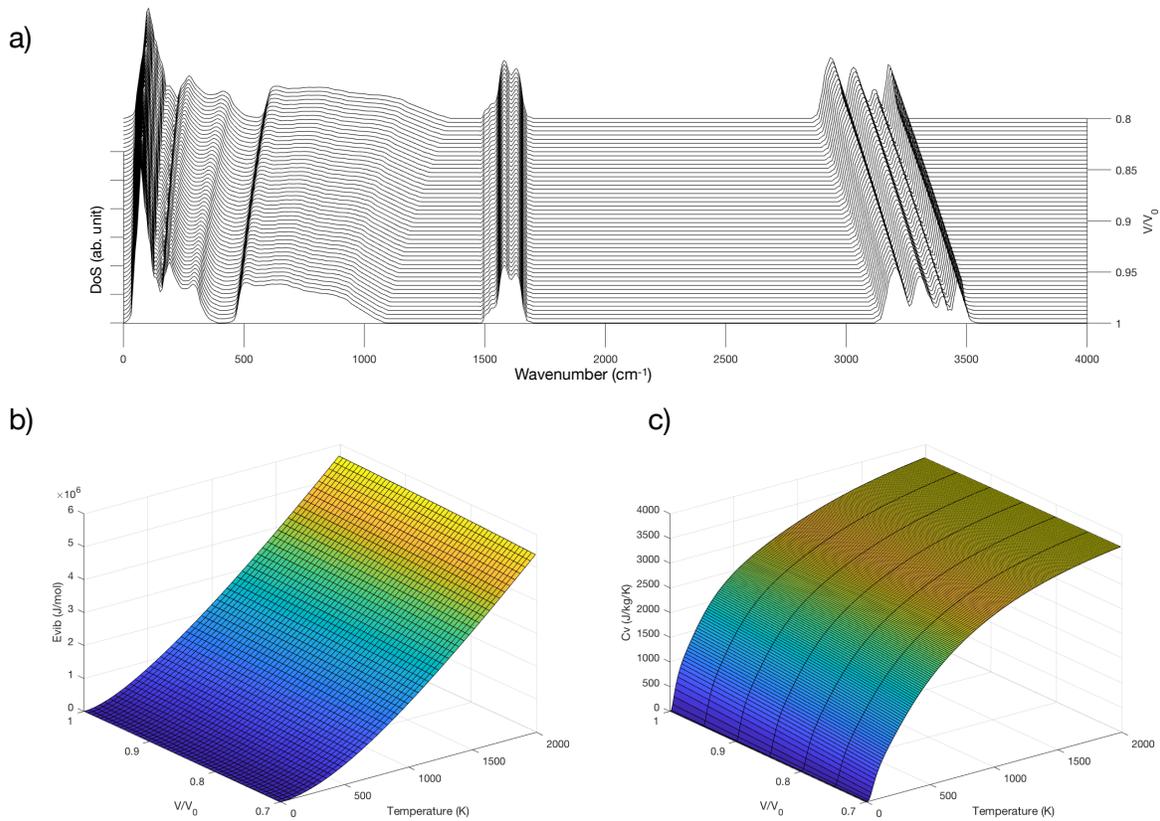

Figure 6 : Thermal properties of Ice V. a) Phonon density of states as a function of frequency and volume compression, b) Vibrational energy $E_{vib}$ as a function of compression and temperature, and c) Isochoric specific heat, $C_v$ as function of compression and temperature.

**Adjustment of model parameters**

For each ice phase the four parameters ($V_o$, $K_o$, $K_o'$, and $\gamma_o$) were adjusted to best fit both diffraction-determined volume measurements and the high-pressure determinations of the adiabatic bulk modulus based on ultrasonic and Brillouin measurements (Gagnon et al. 1988, 1990; Tulk, Kiefte, et al. 1997; Tulk, Gagnon, et al. 1997; Shimizu et al. 1996; Shaw 1986).

In the IAPWS-95 representation of water (Wagner and Pruss 2002), internal energy and entropy are set to zero at its triple point. This convention was also adapted recently for the Gibbs energy representation of water by Bollengier et al. (2019), which is a more accurate representation than IAPWS-95 at high pressures and in the supercooled regime. Here the reference energy $G_o$ of each ice phase was then set to be internally consistent with Bollengier et al. (2019) values by matching the Gibbs energy of the ice phase with water at each lower pressure triple point on the melt line (e.g. ice Ih-III-liquid triple point for ice III).

The entropy of the ices, important in determining the Clapeyron slopes, is the sum of a small "configurational" entropy associated with proton disorder (as estimated by Herrero and Ramírez 2014) and a larger contribution from vibrational entropy. The later were modified from initial estimates by small adjustments, within measurement and calculational uncertainties, to the frequencies of translational and libration modes (by factors of 1.01 to 1.07). These adjustments allowed a better match of melting points over the entire melting range of pressures for each ice phase.

## Acknowledgements

The authors acknowledge the financial support provided by the NASA Postdoctoral Program fellowship awarded to BJ, by the NASA Solar System Workings Grant 80NSSC17K0775 and by the Icy Worlds node of NASA's Astrobiology Institute (08-NAI5-0021). Synchrotron radiation experiments were conducted under beamtime granted to BJ at the ID-15B beamline (proposal number: ES640) at the European Synchrotron Radiation Facility, Grenoble, France. JO financial support was provided by the Department of Earth and Space Science of the University of Washington, Seattle, USA.
The authors would like to thank Dr. Evan Abramson, Dr. Olivier Bollengier, Dr. Tiziana Boffa Ballaran, Dr. Steve Vance, Dr. Gabriel Tobie and Dr. Christophe Sotin for the fruitful scientific discussions that greatly helped enhancing the quality of this work.

## Author contributions

BJ is the leading author of this manuscript. BJ and JMB wrote the manuscript with input from all authors. BJ, AP, SP, IC, and JO operated the high-pressure X-ray diffraction experiments at the ID-15B beamline of the European Synchrotron Radiation Facility, Grenoble France with the scientific and technical support of FC, GG and MH. AP and IC analyzed the X-ray diffraction data. BJ and JMB developed and produced the Gibbs energy equations of state for the ice polymorphs, based on the LBF framework developed by JMB. The opensource SeaFreeze code was written in MATLAB and Python by BJ, PE and JMB.

## Competing interests

The authors declare no competing interests

## Materials & Correspondence.

Correspondence and requests for materials should be addressed to BJ (bjournau@uw.edu)

## Data Availability

The datasets generated and analyzed during the current study are available from the corresponding author on reasonable request.

## Code Availability

The SeaFreeze open-source code is available as supplementary materials and through GitHub (https://github.com/Bjournaux/SeaFreeze)

# Tables

Table 1: Experimental PVT data. Typical uncertainties are estimated to be 30 Mpa in pressure, 0.5K in temperature and $5 \cdot 10^{-3}$ Å$^3$ in volume.

| | Ice III | | | Ice III (continued) | |
|---|---|---|---|---|---|
| Pressure (MPa) | Temperature (K) | Volume (Å$^3$) | Pressure (MPa) | Temperature (K) | Volume (Å$^3$) |
| 260 | 250.5 | 309.168 | 410 | 245.5 | 304.460 |
| 350 | 250.5 | 306.188 | 410 | 245.5 | 304.390 |
| 480 | 250.5 | 303.449 | 390 | 240.5 | 304.424 |
| 480 | 250.5 | 303.200 | 390 | 240.5 | 304.790 |
| 400 | 250.5 | 304.830 | 340 | 240.5 | 305.790 |
| 320 | 250.5 | 307.266 | 340 | 240.5 | 305.610 |
| 289 | 250.5 | 308.127 | 320 | 240.5 | 306.319 |
| 220 | 250.5 | 310.610 | 290 | 240.5 | 306.909 |
| 260 | 245.5 | 308.534 | 260 | 240.5 | 308.303 |
| 260 | 245.5 | 308.630 | 260 | 240.5 | 308.110 |
| 300 | 245.5 | 307.750 | 220 | 240.5 | 309.277 |
| 340 | 245.5 | 306.526 | 210 | 240.5 | 310.157 |
| 340 | 245.5 | 306.620 | 230 | 252.5 | 310.055 |

| | Ice V | | | Ice V (continued) | |
|---|---|---|---|---|---|
| Pressure (MPa) | Temperature (K) | Volume (Å$^3$) | Pressure (MPa) | Temperature (K) | Volume (Å$^3$) |
| 689 | 242.3 | 654.867 | 566 | 252.3 | 660.269 |
| 562 | 242.3 | 660.141 | 578 | 249.3 | 660.479 |
| 531 | 242.3 | 662.162 | 581 | 243.3 | 659.901 |
| 496 | 242.3 | 663.845 | 583 | 237.3 | 659.206 |
| 421 | 242.3 | 667.471 | 581 | 232.3 | 659.638 |
| 506 | 262.3 | 665.551 | 603 | 222.3 | 658.473 |
| 538 | 262.3 | 663.699 | 647 | 222.3 | 656.677 |
| 548 | 262.3 | 663.103 | 704 | 222.3 | 654.532 |
| 500 | 252.3 | 664.129 | 779 | 222.3 | 651.850 |

| | Ice VI | | | Ice VI (continued) | |
|---|---|---|---|---|---|
| Pressure (MPa) | Temperature (K) | Volume (Å$^3$) | Pressure (MPa) | Temperature (K) | Volume (Å$^3$) |
| 1007 | 242.3 | 218.895 | 889 | 262.3 | 220.434 |
| 1537 | 242.3 | 214.056 | 1041 | 262.3 | 218.331 |
| 1351 | 242.3 | 215.042 | 1216 | 262.3 | 216.399 |
| 1217 | 242.3 | 216.174 | 1452 | 262.3 | 214.312 |
| 1094 | 242.3 | 217.336 | 476 | 262.3 | 224.652 |
| 913 | 242.3 | 219.730 | 726 | 262.3 | 221.252 |
| 746 | 242.3 | 221.642 | 903 | 262.3 | 219.731 |

| | 706 | 262.3 | 222.595 |
|---|---|---|---|

Table 2 : Ice polymorphs constructed Density of States details.

| Ice Polymorphs<br>*Mode family* | Type and conditions | Reference | $\gamma_i$ | $\mathcal{N}_i$ |
|---|---|---|---|---|
| **Ice III** | | | | |
| *Translation* | Computed (Q-TIP4P/F) / P = 0 Mpa, $V_{ref}$ = 24.99 Å /molec. | Ramírez et al. (2012) | 1.3 | 3 |
| *Libration* | Computed (Q-TIP4P/F) / P = 0 Mpa, $V_{ref}$ = 24.99 Å /molec. | Ramírez et al. (2012) | 0.25 | 3 |
| *Bending* | Computed (Q-TIP4P/F) / P = 0 Mpa, $V_{ref}$ = 24.99 Å /molec. | Ramírez et al. (2012) | 0.05 | 1 |
| *Stretching* | Computed (Q-TIP4P/F) / P = 0 Mpa, $V_{ref}$ = 24.99 Å /molec. | Ramírez et al. (2012) | -0.1 | 2 |
| **Ice V** | | | | |
| *Translation* | Measured INS / P = 20 mbar, T < 15 K | (Li 1996) | 1.6 | 3 |
| *Libration* | Measured INS P = 20 mbar, T < 15 K | (Li 1996) | *1 | 3 |
| *Bending* | Computed / P = 0 Mpa, $V_{ref}$ = 24.27 Å /molec | (Jenkins and Morrison 2001) | *0 | 1 |
| *Stretching* | Computed / P = 0 Mpa, $V_{ref}$ = 24.27 Å /molec | (Jenkins and Morrison 2001) | -0.4 | 2 |
| **Ice VI** | | | | |
| *Translation* | Measured INS / P = 20 mbar, T < 15 K | (Li 1996) | 2.5 | 3 |
| *Libration* | Measured INS / P = 20 mbar, T < 15 K | (Li 1996) | *1 | 3 |
| *Bending* | Computed / P = 0 Mpa, $V_{ref}$ = 20.88 Å /molec | (Jenkins and Morrison 2001) | *0 | 1 |
| *Stretching* | Computed / P = 0 Mpa, $V_{ref}$ = 20.88 Å /molec | (Jenkins and Morrison 2001) | -0.44 | 2 |

Table 3 : Ice polymorphs Mie-Grüneisen Equations of States fit parameters.

| Phase | $V_0$ (m³/kg) | $V_0$ (cm³/mol) | $K_0$ (GPa) | $K'_0$ | $\gamma$ | $q$ | Reference |
|---|---|---|---|---|---|---|---|
| Ice III | 8.595·10⁻⁴ | 15.49(5) | 9.9(3) | 6 | 1.0 | 1 | this study |
| Ice V | 8.035·10⁻⁴ | 14.48(5) | 13.2(3) | 6 | 1.1 | 1 | this study |
| Ice VI | 7.562·10⁻⁴ | 13.62(2) | 15.2(3) | 6.5 | 1.4 | 1 | this study |

Table 4: Values for the triple point coordinates, constrains to the Gibbs energy surface LBF representation. Configurational entropies $S_c$, are from Herrero and Ramírez (2014). Gibbs energies are given in J·kg⁻¹ and entropies in J·K⁻¹·kg⁻¹

| | Triple point | | | | Constrains values | | LBF Gibbs representation values | | |
|---|---|---|---|---|---|---|---|---|---|
| Phase | Triple point phases | $P_{TP}$ (MPa) | $T_{TP}$ (K) | Ref (TP) | $G_w(P_{TP}, T_{TP})$ (Bollengier et al. 2019) | $S_c$ | $G_o(P_{TP}, T_{TP})$ | $S_o(P_{TP}, T_{TP})$ | $S_c$ |
| Ice III | Ih-III-L | 208.566 | 251.165 | Wagner et al. (2011) | 1.9528·10⁵ | 189.99 | 4.3452·10⁵ | 3326.4 | 189.99 |
| Ice V | III-V-L | 350.1 | 256.164 | Wagner et al. (2011) | 3.2394·10⁵ | 190.68 | 5.7396·10⁵ | 3327.7 | 190.68 |
| Ice VI | V-VI-L | 632.4 | 273.31 | Wagner et al. (2011) | 5.7096·10⁵ | 194.47 | 8.4519·10⁵ | 3346.5 | 194.47 |